\documentclass[review]{elsarticle}
\usepackage{graphicx}
\pdfoutput=1

\newtheorem{theorem}{Theorem}
\newtheorem{lemma}[theorem]{Lemma}
\newtheorem{corollary}[theorem]{Corollary}

\newcommand{\Oh}[1]
    {\ensuremath{\mathcal{O}\!\left( {#1} \right)}}
\newcommand{\occ}
    {\ensuremath{\mathrm{occ}}}

\begin{document}

\begin{frontmatter}

\title{A Faster Grammar-Based Self-Index}

\author[aalto]{Travis Gagie}
\author[mpi]{Pawe\l\ Gawrychowski}
\author[helsinki]{Juha K\"arkk\"ainen}
\author[chile]{\\Yakov Nekrich}
\author[helsinki]{Simon J. Puglisi}
\address[aalto]{Aalto University}
\address[mpi]{Max Planck Institute}
\address[helsinki]{University of Helsinki}
\address[chile]{University of Chile}

\begin{abstract}
To store and search genomic databases efficiently, researchers have recently started building compressed self-indexes based on grammars.  In this paper we show how, given a straight-line program with $r$ rules for a string \(S [1..n]\) whose LZ77 parse consists of $z$ phrases, we can store a self-index for $S$ in $\Oh{r + z \log \log n}$ space such that, given a pattern \(P [1..m]\), we can list the $\occ$ occurrences of $P$ in $S$ in $\Oh{m^2 + \occ \log \log n}$ time.  If the straight-line program is balanced and we accept a small probability of building a faulty index, then we can reduce the $\Oh{m^2}$ term to $\Oh{m \log m}$.  All previous self-indexes are larger or slower in the worst case.
\end{abstract}

\begin{keyword}
compressed self-indexes \sep grammar-based compression \sep Lempel-Ziv compression
\end{keyword}

\end{frontmatter}

\section{Introduction} \label{sec:intro}

With the advance of DNA-sequencing technologies comes the problem of how to store many individuals' genomes compactly but such that we can search them quickly.  Any two human genomes are 99.9\% the same, but compressed self-indexes based on compressed suffix arrays, the Burrows-Wheeler Transform or LZ78 (see~\cite{NM07} for a survey) do not take full advantage of this similarity.  Researchers have recently started building self-indexes based on context-free grammars (CFGs) and LZ77~\cite{ZL77}, which better compress highly repetitive strings.  A compressed self-index stores a string \(S [1..n]\) in compressed form such that, first, given a position $i$ and a length $\ell$, we can quickly extract \(S [i..i + \ell - 1]\) and, second, given a pattern \(P [1..m]\), we can quickly list the $\occ$ occurrences of $P$ in $S$.

\begin{figure}[t]
\begin{center}
\resizebox{60ex}{!}
{\begin{tabular}{l@{\hspace{5ex}}r}
\parbox{20ex}
{\vspace{-25ex}
\begin{eqnarray*}
X_7 & \rightarrow & X_6 X_5\\
X_6 & \rightarrow & X_5 X_4\\
X_5 & \rightarrow & X_4 X_3\\
X_4 & \rightarrow & X_3 X_2\\
X_3 & \rightarrow & X_2 X_1\\
X_2 & \rightarrow & a\\
X_1 & \rightarrow & b
\end{eqnarray*}} &
\includegraphics[width=50ex]{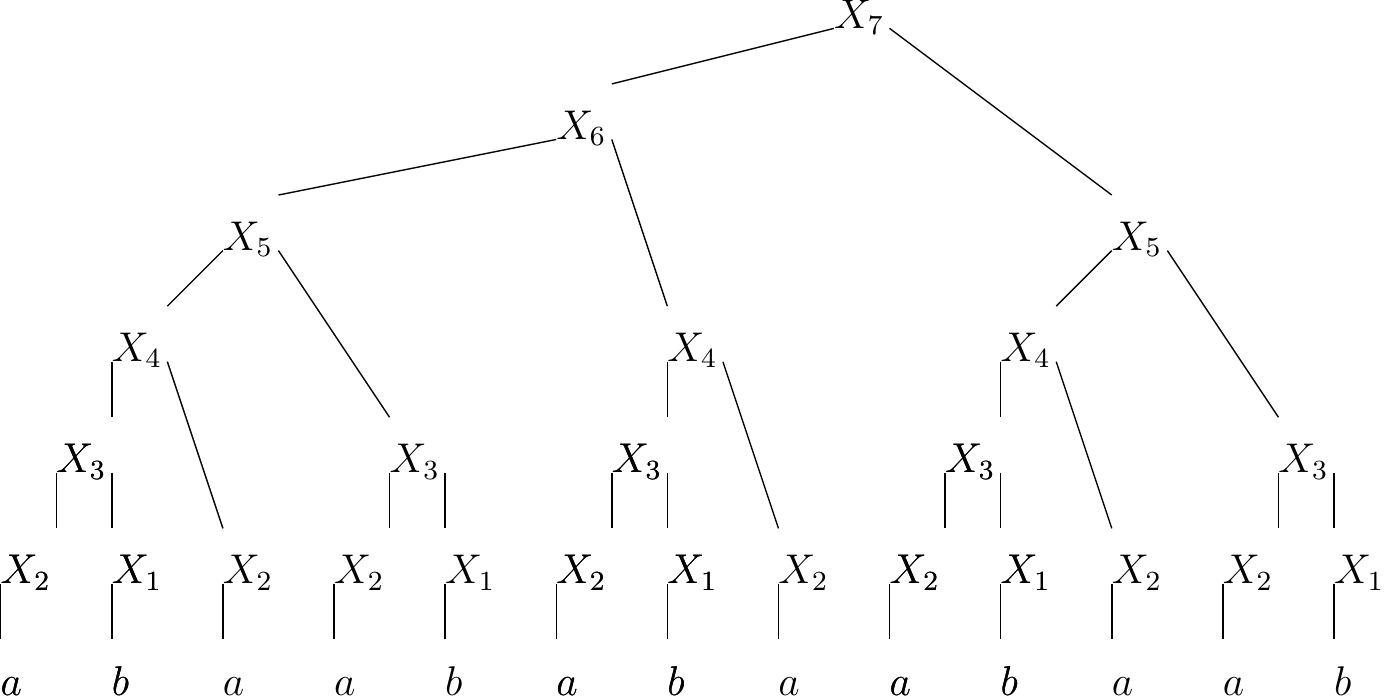}
\end{tabular}}
\caption{A balanced SLP for \(abaababaabaab\) (left) and the corresponding parse tree (right).}
\label{fig:slp}
\end{center}
\end{figure}

Claude and Navarro~\cite{CN11a} gave the first compressed self-index based on grammars or, more precisely, straight-line programs (SLPs).  An SLP is a context-free grammar (CFG) in Chomsky normal form that generates only one string.  Figure~\ref{fig:slp} shows an example.  They showed how, given an SLP with $r$ rules for a string $S$, we can build a self-index that takes $\Oh{r}$ space and supports extraction in $\Oh{(\ell + h) \log r}$ time and pattern matching in $\Oh{(m (m + h) + h\,\occ) \log r}$ time, respectively, where $h$ is the height of the parse tree.  Our model is the word RAM with $\Theta (\log n)$-bit words; except where stated otherwise, by $\log$ we mean $\log_2$ and we measure space in words.  The same authors~\cite{CN11b} recently gave a self-index that has better time bounds and can be based on any CFG generating $S$ and only $S$.  Specifically, they showed how, given such a CFG with $r'$ distinct terminal and non-terminal symbols and $R$ symbols on the righthand sides of the rules, we can build a self-index that takes $\Oh{R}$ space and supports extraction in $\Oh{\ell + h \log (R / h)}$ time and pattern matching in $\Oh{m^2 \log (\log n / \log r') + \occ \log r'}$ time.

If we are not concerned about the constant coefficient in the space bound, we can improve Claude and Navarro's time bound for extraction.  Calculation shows that \(h \log (R / h) \geq \log n\).  Given a CFG generating $S$ and only $S$ with $R$ symbols on the righthand sides of the rules, we can turn it into an SLP with $\Oh{R}$ rules (although the number of distinct symbols and the height of the parse tree can each increase by a factor of $\Oh{\log n}$).  Bille et al.~\cite{BLRSSW11} showed how we can store such an SLP in $\Oh{R}$ space and support extraction in $\Oh{\ell + \log n}$ time.  Combining their result with Claude and Navarro's improved one, we obtain an index that still takes $\Oh{R}$ space and $\Oh{m^2 \log (\log (n) / \log r') + \occ \log r'}$ time for pattern matching but only $\Oh{\ell + \log n}$ time for extraction.

In this paper we show how, given an SLP for $S$ with $r$ rules, we can build a self-index that takes $\Oh{r + z \log \log n}$ space, where $z$ is the number of phrases in the LZ77 parse of $S$, and supports extraction in $\Oh{\ell + \log n}$ time and pattern matching in $\Oh{m^2 + \occ \log \log n}$ time.  Therefore, by the observations above, given a CFG generating $S$ and only $S$ with $R$ symbols on the righthand sides of the rules, we can build an index with the same time bounds that takes $\Oh{R + z \log \log n}$ space.

If we are given a balanced SLP for $S$ --- i.e., one for which the parse tree is height- or weight-balanced~\cite{CLRS01} --- and we accept a small probability of building a faulty index, then we do not need Bille et al.'s result to extract in $\Oh{\ell + \log n}$ time and we can reduce the time bound for pattern matching to $\Oh{m \log m + \occ \log \log n}$.  Rytter~\cite{Ryt03} showed how we can build such an SLP with $\Oh{z \log (n / z)}$ rules, and proved that no SLP for $S$ has fewer than $z$ rules.  His algorithm still has the best known approximation ratio even when the SLP need not be balanced, but performs badly in practice.  Recently, however, Maruyama, Sakamoto and Takeda~\cite{MST12} gave a practical online algorithm that produces a balanced SLP with $\Oh{z \log^2 n}$ rules.  In other words, requiring the SLP to be balanced is a reasonable restriction both in theory and in practice.

Table~\ref{tab:bounds} summarizes Claude and Navarro's bounds and our own.  Since all the self-indexes mentioned can be made to support extraction in $\Oh{\ell + \log n}$ time without increasing their space usage by more than a constant factor, we do not include this bound in the table.  As noted above, given a CFG generating $S$ and only $S$ with $R$ symbols on the righthand sides of the rules, we can turn it into an SLP with $\Oh{R}$ rules, so our first result is as general as Claude and Navarro's; the $r$ in the second row of the table can be replaced by $R$.  By Rytter's result, we can assume \(z \leq r = \Oh{z \log (n / z)}\).

\begin{table}[t]
\begin{center}
\caption{Claude and Navarro's bounds and our own.  In the first row, $R$ is the number of symbols on the righthand sides of the rules in a given CFG generating $S$ and only $S$, and $r'$ is the number of distinct terminal and non-terminal symbols in that CFG.  In the second and third rows, $r$ is the number of rules in a given SLP for $S$ --- which must be balanced in the third row --- and $z$ is the number of phrases in the LZ77 parse of $S$.}
\label{tab:bounds}
\vspace{2ex}
\begin{tabular}{c@{\hspace{2ex}}|@{\hspace{2ex}}c@{\hspace{4ex}}c}
source & space & search time\\
\hline &&\\[-1ex]
\cite{CN11b} &
$\Oh{R}$ &
$\Oh{m^2 \log \left( \frac{\log n}{\log r'} \right) + \occ \log r'}$ \\[2ex]
Theorem~\ref{thm:unbalanced} &
$\Oh{r + z \log \log n}$ &
$\Oh{m^2 + \occ \log \log n}$ \\[1ex]
Theorem~\ref{thm:balanced} &
$\Oh{r + z \log \log n}$ &
$\Oh{m \log m + \occ \log \log n}$
\end{tabular}
\end{center}
\end{table}

There are other self-indexes optimized for highly repetitive strings but comparing ours against them directly is difficult.  For example, Do et al.'s~\cite{DJSS12} space bound is in terms of the number of phrases in a new variant of the LZ77 parse~\cite{KPZ10}, which can be much larger than $z$; Huang et al.'s~\cite{HLSTY10} is bounded in terms of the number and length of common and distinct regions in the text; Maruyama et al.'s~\cite{MNKS11} time bound for pattern matching depends on ``the number of occurrences of a maximal common subtree in [the edit-sensitive parse] trees of $P$ and $S$''; Kreft and Navarro's~\cite{KN12} time bound depends on the depth of nesting in the LZ77 parse.

We still use many ideas from Kreft and Navarro's work, which we describe in Section~\ref{sec:KN12}.  In Section~\ref{sec:unbalanced} we show how, given an SLP for $S$ with $r$ rules, we can build a self-index that takes $\Oh{r + z \log \log n}$ space and supports extraction in $\Oh{\ell + \log n}$ time and pattern matching in $\Oh{m^2 + \occ \log \log n}$ time.  We also show how, with the same self-index, in $\Oh{m^2 \log \log n}$ time we can compute all cyclic shifts and maximal substrings of $P$ that occur in $S$.  In Section~\ref{sec:unbalanced} we show how, if the SLP is balanced and we accept a small probability of building a faulty index, then we can reduce the time bound for pattern matching to $\Oh{m \log m + \occ \log \log n}$.  Finally, in Section~\ref{sec:future} we discuss directions for future work.

\section{Kreft and Navarro's Self-Index} \label{sec:KN12}

\begin{figure}[t]
\begin{center}
\resizebox{70ex}{!}{\includegraphics{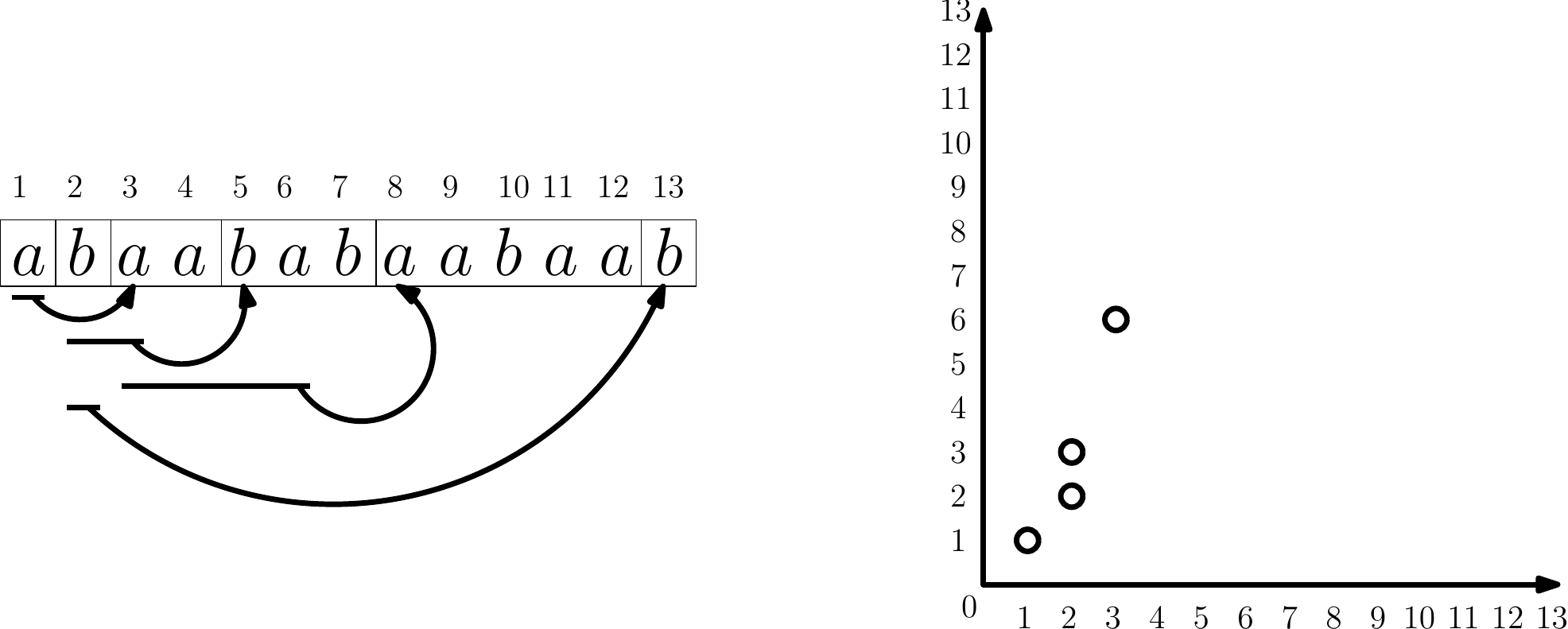}}
\caption{The LZ77 parse of ``\(abaababaabaab\)'' (left) and the locations of the phrase sources plotted as points on a grid (right).  In the parse, horizontal lines indicate phrases' sources, with arrows leading to the boxes containing the phrases themselves.  On the grid, a point's horizontal coordinate is where the corresponding source starts, and its vertical coordinate is where the source ends.  Notice that a phrase source \(S [i..j]\) covers a substring \(S [i'..i' + m - 1]\) if and only if the point \((i, j)\) is above and to the left of the point \((i', i' + m - 1)\).}
\label{fig:parse}
\end{center}
\end{figure}

The LZ77 compression algorithm works by parsing $S$ from left to right into $z$ phrases: after parsing \(S [1..i - 1]\), it finds the longest prefix \(S [i..j - 1]\) of \(S [i..m]\) that has occurred before and selects \(S [i..j]\) as the next phrase.  If \(j = 1\) then the phrases consists only of the first occurrence of a character; otherwise, the leftmost occurrence of \(S [i..j - 1]\) is called the phrase's source.  Figure~\ref{fig:parse} shows an example.

Kreft and Navarro~\cite{KN12} gave the first (and, so far, only) compressed self-index based on LZ77.  Their index takes \(\Oh{z \log n} + o (n)\) bits and supports extraction in $\Oh{\ell d}$ time and pattern matching in $\Oh{m^2 d + (m + \occ) \log z}$ time, where \(d \leq z\) is the depth of nesting in the parse.  They considered only the non-self-referential version of LZ77, so \(z \geq \log n\); for ease of comparison, we do the same.  They also gave a variant of LZ77 called LZ-End, with which they can reduce the extraction time to $\Oh{\ell + d}$.  Although they showed that LZ-End performs well in practice, however, they were unable to bound the worst-case size of the LZ-End parse in terms of $z$.

\begin{figure}[t]
\begin{center}
\resizebox{70ex}{!}{\includegraphics{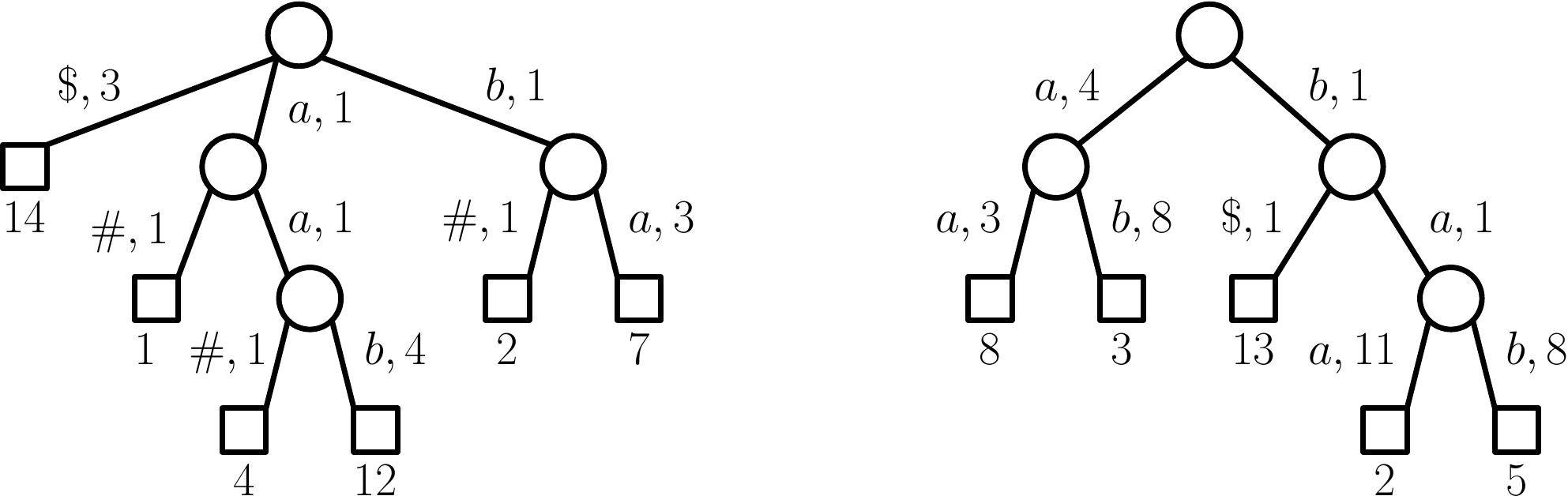}}
\caption{The Patricia trees for the reversed phrases (left) and suffixes starting at phrase boundaries (right) in the LZ77 parse of ``\(abaababaabaab\)''.}
\label{fig:Patricia}
\end{center}
\end{figure}

Kreft and Navarro start by building two Patricia trees, one for the reverses of the phrases in the LZ77 parse and the other for the suffixes of $S$ that start at phrase boundaries.  A Patricia tree~\cite{Mor68} is a compacted trie for substrings of a stored string, in which we store only the first character and length of each edge label; the leaves store pointers into the string itself such that, after finishing a search at a node in the tree, we can verify that the node's path label matches the string we seek.  The total size of the two Patricia trees is $\Oh{z}$.  Since Kreft and Navarro store $S$ in compressed form, they extract nodes' path labels in order to verify them.  For example, if \(S = abaababaabaab\), then the reverses of the phrases are shown below on the left with the phrase numbers, in order by phrase number on the left and in lexicographic order on the right; the suffixes starting at phrase boundaries are shown on the right.  When building the Patricia trees, we treat $S$ as ending with a special character $\$$ lexicographically less than any in the alphabet, and each reversed phrase as ending with another special character $\#$.  Figure~\ref{fig:Patricia} shows the Patricia trees for this example.
\begin{center}
\begin{tabular}{rl@{\hspace{3ex}}rl@{\hspace{8ex}}rl@{\hspace{3ex}}rl}
1) & \(a\#\) & 6) & \(\$b\#\) & 1) & \(baababaabaab\$\) & 6) & $\epsilon$\\
2) & \(b\#\) & 1) & \(a\#\) & 2) & \(aababaabaab\$\) & 4) & \(aabaab\$\)\\
3) & \(aa\#\) & 3) & \(aa\#\) & 3) & \(babaabaab\$\) & 2) & \(aababaabaab\$\)\\
4) & \(bab\#\) & 5) & \(aabaa\#\) & 4) & \(aabaab\$\) & 5) & $b$\\
5) & \(aabaa\#\) & 2) & \(b\#\) & 5) & \(b\$\) & 1) & \(baababaabaab\$\)\\
6) & \(\$b\#\) & 4) & \(bab\#\) & 6) & $\epsilon$ & 3) & \(babaabaab\$\)
\end{tabular}
\end{center}

\begin{figure}[t]
\begin{center}
\resizebox{50ex}{!}{\includegraphics{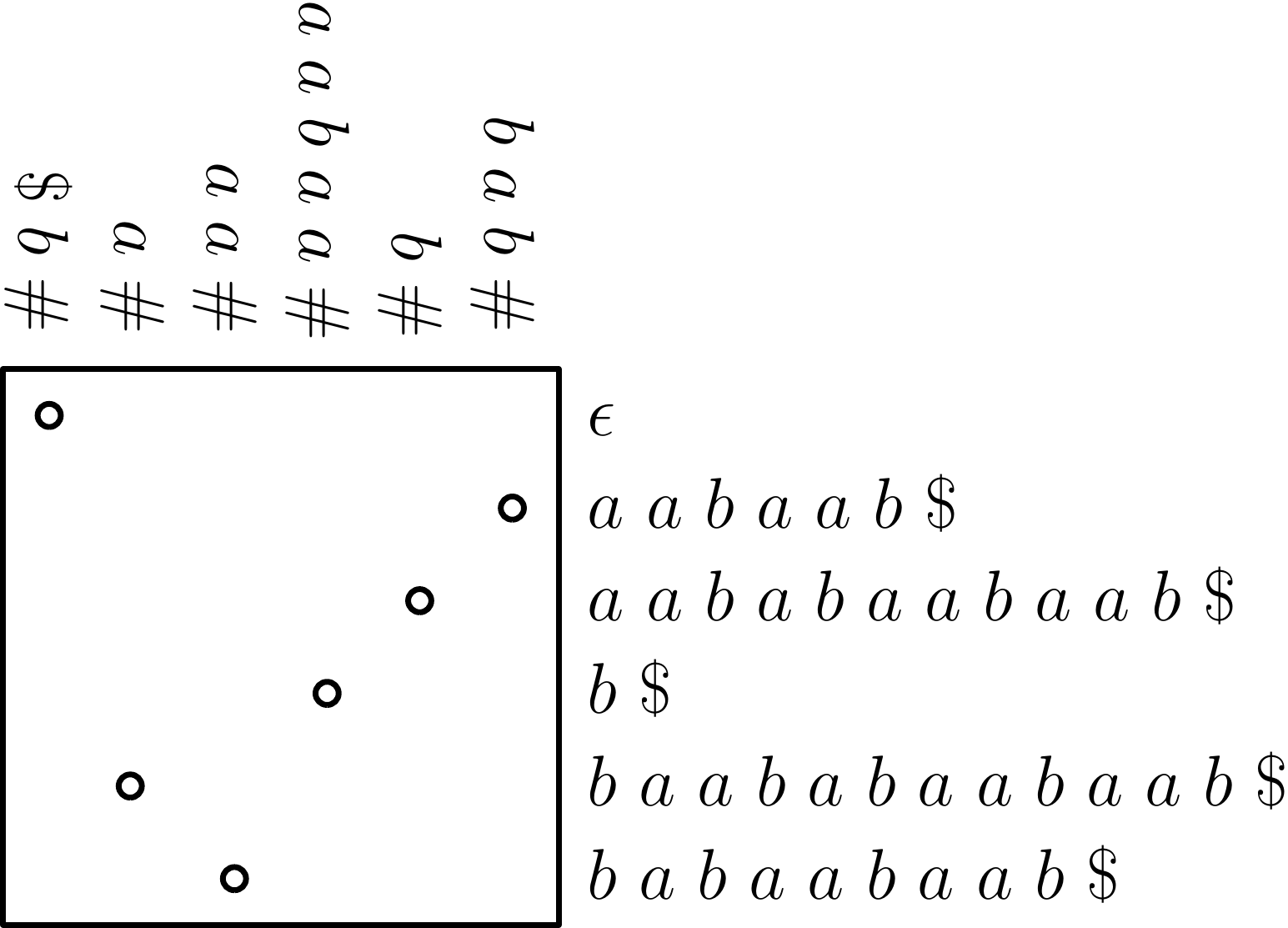}}
\caption{A grid showing how, in the LZ77 parse of ``\(abaababaabaab\)'', reversed phrases precede suffixes starting at phrase boundaries.}
\label{fig:grid}
\end{center}
\end{figure}

Their next component is a data structure for four-sided range reporting on a \(z \times z\) grid storing $z$ points, with each point \((i, j)\) indicating that the lexicographically $i$th reversed phrase is followed in $S$ by the lexicographically $j$th suffix starting at a phrase boundary.  Figure~\ref{fig:grid} shows the grid for our running example \(S = abaababaabaab\).  Kreft and Navarro use a wavelet tree, which takes $\Oh{z}$ space and answers queries in $\Oh{(p + 1) \log z}$ time, where $p$ is the number of points reported~\cite{Nav12}.  Many other data structures are known for this problem, however, with different time-space tradeoffs.

Their final component is (essentially) a data structure for two-sided range reporting on an \(n \times n\) grid storing at most \(z - 1\) points, with each point \((i, j)\) indicating that \(S [i..j]\) is a phrase's source.  The grid for \(S = abaababaabaab\) is shown beside the LZ77 parse in Figure~\ref{fig:parse}.  They implement this data structure with a compressed bitvector (as a predecessor data structure) and a range-minimum data structure, which take \(\Oh{z \log n} + o (n)\) bits of space and answer queries in $\Oh{p + 1}$ time, where $p$ is again the number of points reported.  Again, however, other time-space tradeoffs are available.

Given a pattern \(P [1..m]\), Kreft and Navarro use the two Patricia trees to find, for \(1 \leq i \leq m\), the lexicographic range of the reverses of phrases ending with \(P [1..i]\), and the lexicographic range of the suffixes starting with \(P [i + 1..m]\) at phrase boundaries.  This takes a total of $\Oh{m^2}$ time to descend the Patricia trees and $\Oh{m^2 d}$ time to extract nodes' path labels.  They then use the wavelet tree to find all the phrase boundaries preceded by \(P [1..i]\) and followed by \(P [i + 1..m]\), which takes a total of $\Oh{(m + \occ) \log z}$ time.  After these steps, they know the locations of all occurrences of $P$ that cross phrase boundaries in $S$, which are called primary occurrences.

An occurrence of $P$ that is completely contained within a phrase is called a secondary occurrence.  By the definition of LZ77, the first occurrence must be primary and any secondary occurrence must be copied from an earlier occurrence.  We can find all secondary occurrences by finding all primary occurrences and then recursively finding all phrase sources that cover occurrences we have already found.  Notice that, if a phrase source \(S [i..j]\) covers an occurrence \(S [i'..i' + m - 1]\), then \(i \leq i'\) and \(j \geq i' + m - 1\), so the point \((i, j)\) is above and to the left of the point \((i', i' + m - 1)\).  It follows that, after finding all primary occurrences of $P$, Kreft and Navarro can find all secondary occurrences in $\Oh{\occ}$ time using one two-sided range reporting per occurrence.  Therefore, their self-index supports pattern matching in a total of $\Oh{m^2 d + (m + \occ) \log z}$ time.

We can use a new data structure by Chan, Larsen and P\v{a}tra\c{s}cu~\cite{CLP11} for four-sided range reporting, instead of a wavelet tree, and a y-fast trie~\cite{Wil83} for predecessor queries, instead of a compressed bitvector.  Calculation shows that Kreft and Navarro's space bound then changes to $\Oh{z \log \log z}$ words and their time bound improves to $\Oh{m^2 d + m \log \log z + occ \log \log n}$.  Bille and G{\o}rtz~\cite{BG11} showed how, by storing one-dimensional range-reporting data structures at each node in the top \(\log \log z\) levels of the Patricia trees, we can eliminate the $\Oh{m \log \log z}$ term: if \(m \leq \log \log z\) then instead of the data structure for four-sided range reporting, we can use the one-dimensional range-reporting data structures, which are faster; otherwise, the $\Oh{m^2}$ term dominates the $\Oh{m \log \log z}$ term anyway.  Thus, by implementing the components differently in Kreft and Navarro's self-index, we obtain one that takes $\Oh{z \log \log z}$ space and supports pattern matching in $\Oh{m^2 d + occ \log \log n}$ time.

If we are given an SLP for $S$ with $r$ rules then we can also combine Bille et al.'s~\cite{BLRSSW11} with our modification of Kreft and Navarro's.  We can use Bille et al.'s data structure for extracting nodes' path labels while pattern matching, so we obtain a self-index that takes $\Oh{r + z \log \log z}$ space and supports extraction in $\Oh{\ell + \log n}$ time and pattern matching in $\Oh{m^2 + m \log n + \occ \log \log n}$ time.  In Section~\ref{sec:unbalanced} we explain how to remove the $\Oh{m \log n}$ term by taking advantage of the fact that, while pattern matching, we extract nodes' path labels only from phrase boundaries.

\section{Self-Indexing with an Unbalanced SLP} \label{sec:unbalanced}

Suppose we are given an SLP for $S$ with $r$ rules and a list of $t$ specified positions from which we want to support linear-time extraction, e.g., from the phrase boundaries in the LZ77 parse.  We can build an instance of Bille et al.'s~\cite{BLRSSW11} data structure and support extraction from any position in $\Oh{\ell + \log n}$ time, where $\ell$ is the length of the substring extracted.  When \(\ell = \Omega (\log n)\) we have \(\Oh{\ell + \log n} = \Oh{\ell}\), i.e., the extraction is linear-time.  Therefore, we need worry only about extracting substrings of length \(o (\log n)\) from around the $t$ specified positions.

Consider each substring that starts \(\log n\) characters to the left of a specified position and ends \(\log n\) characters to the right of that position.  By the definition of LZ77, the first occurrence of that substring crosses a phrase boundary.  If we store a pointer to the first occurrence of each such substring, which takes $\Oh{t}$ space, then we need worry only about extracting substrings of length \(o (\log n)\) from around the phrase boundaries.  Now consider the string \(S' [1..n']\) obtained from $S$ by removing any character at distance more than \(\log n\) from the nearest phrase boundary.  Notice that $S'$ can be parsed into $\Oh{z}$ substrings, each of which
\begin{itemize}
\item occurs in $S$,
\item has length at most \(\log n\),
\item is either a single character or does not touch a phrase boundary in the LZ77 parse of $S$.
\end{itemize}
We claim that any such substring \(S' [i..j]\) is split between at most 2 phrases in the LZ77 parse of $S'$.  To see why, consider that the first copy of \(S' [i..j]\) in $S$ must touch a phrase boundary and is completely within distance \(\log n\) of that phrase boundary, so it remains intact in $S'$.  Therefore, either \(S' [i..j]\) is a single character --- which is obviously contained within only 1 phrase in the LZ77 parse of $S'$ --- or \(S' [i..j]\) is not the first occurrence of that substring in $S'$.  It follows that the LZ77 parse of $S'$ consists of $\Oh{z}$ phrases.  Clearly \(n' = \Oh{z \log n}\), so we can apply Rytter's algorithm to build a balanced SLP for $S'$ that has \(r' = \Oh{z \log (n' / z)} = \Oh{z \log \log n}\) rules.  Since this SLP is balanced, its parse tree has height $\Oh{\log z + \log \log n}$ and so we can store it in \(\Oh{r'} = \Oh{z \log \log n}\) space and support extraction from any position in $S'$ in \(\Oh{\ell + \log n'} = \Oh{\ell + \log z}\) time.

We now have a data structure that takes $\Oh{r + t + z \log \log n}$ space and supports extraction from any position in $S$ in $\Oh{\ell + \log n}$ time and extraction from the $t$ specified positions in $\Oh{\ell + \log z}$ time.  If we choose the specified positions to be the phrase boundaries in the LZ77 parse of $S$, then we can combine it with our modification of Kreft and Navarro's index from Section~\ref{sec:KN12} and obtain a self-index that takes $\Oh{r + z \log \log n}$ space and supports extraction in $\Oh{\ell + \log n}$ time and pattern matching in $\Oh{m^2 + m \log z + \occ \log \log n}$ time.  We next eliminate the $\Oh{m \log z}$ term by taking advantage of the fact that the SLP for $S'$ is balanced.

As noted in Section~\ref{sec:intro}, an SLP is balanced if the corresponding parse tree is height- or weight-balanced.  Suppose we are given a position $i$ in $S'$ and a bound $L$ and asked to add $\Oh{1}$ space to our balanced SLP for $S'$ such that we can support extraction of any substring \(S' [i..i + \ell - 1]\) with \(\ell \leq L\) in $\Oh{\ell + \log L}$ time.  Supporting extraction of any substring \(S' [i - \ell + 1..i]\) in $\Oh{\ell + \log L}$ time is symmetric.  We find the lowest common ancestor $u$ of the $i$th and \((i + L)\)th leaves of the parse tree $T$ for $S'$.  We then find the deepest node $v$ in $u$'s left subtree such that $v$'s subtree contains the $i$th leaf of $T$, and the deepest node $w$ in $u$'s right subtree such that $w$'s subtree contains the \((i + L)\)th leaf.  Since our SLP for $S'$ is balanced, $v$ and $w$ have height $\Oh{\log L}$.  To see why, consider that the ancestors of the rightmost leaf in $u$'s subtree and of the leftmost leaf in its right subtree, have exponentially many leaves in their height.

\begin{figure}[t]
\begin{center}
\resizebox{50ex}{!}{\includegraphics{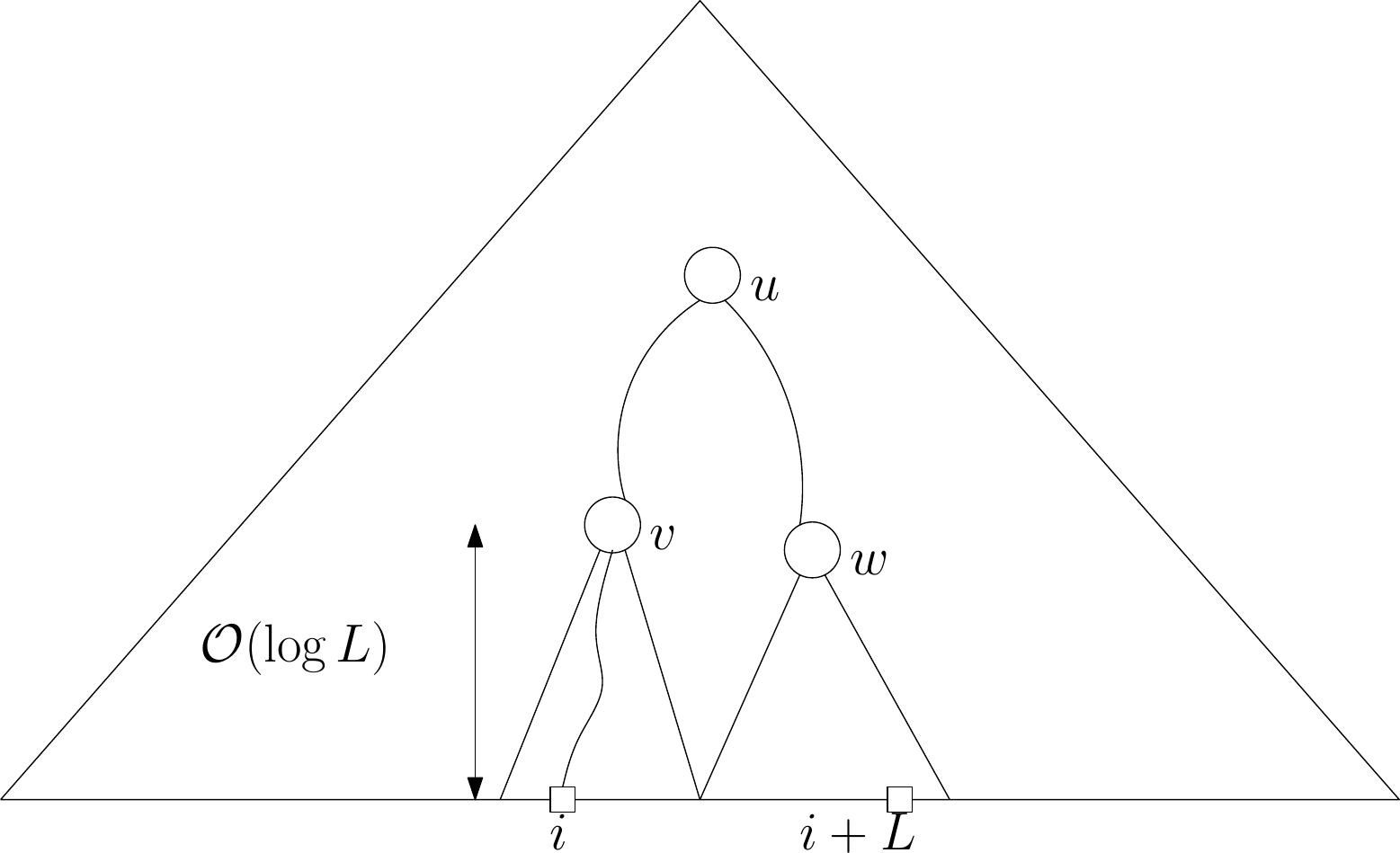}}
\caption{To support fast extraction from position $i$, we store the non-terminals at $v$ and $w$ and the path from $v$ to the $i$th leaf in the parse tree.}
\label{fig:bookmark}
\end{center}
\end{figure}

Without loss of generality, assume $v$'s subtree contains the $i$th leaf.  We store the non-terminals at $v$ and $w$ in $\Oh{\log r'}$ bits, and $\Oh{\log L}$ bits indicating the path from $v$ to the $i$th leaf; together these take $\Oh{1}$ words.  Figure~\ref{fig:bookmark} shows an example.  We can view the symbols at nodes of $T$ as pointers to those nodes and use the rules of the grammar to navigate in the tree.  To extract \(S' [i..i + \ell - 1]\), we start at $v$, descend to the $i$th leaf in $T$, and then traverse the leaves to the right until we have either reached the \((i + \ell - 1)\)st leaf in $T$ or the rightmost leaf in $v$'s subtree; in the latter case, we perform a depth-first traversal of $w$'s subtree until we reach the \((i + \ell - 1)\)st leaf in $T$.  During both traversals we output the terminal symbol at each leaf when we visit it.  If we store the size of each non-terminal's expansion (i.e., the number of leaves in the corresponding subtree of the parse tree) then, after descending from $v$ to the $i$th leaf, in $\Oh{\log L}$ time we can compute a list of $\Oh{\log \ell}$ terminal and non-terminal symbols such that the concatenation of their expansions is \(S' [i..i + \ell - 1]\).  This operation will prove useful in Section~\ref{sec:balanced}.

Since we can extract any substring \(S [i..i + \ell - 1]\) in $\Oh{\ell + \log L}$ time and extracting any substring \(S' [i - \ell + 1..i]\) in $\Oh{\ell + \log L}$ time is symmetric, we can extract any substring of length $\ell$ that crosses position $i$ in $S'$ in $\Oh{\ell + \log L}$ time.  We can already extract any substring in $\Oh{\ell + \log n'}$ time, so we first choose \(L = \log n'\) and store $\Oh{1}$ words to be able to extract any substring that crosses position $i$ in $\Oh{\ell + \log \log n'}$ time.  We then choose \(L = \log \log n'\) and store another $\Oh{1}$ words to be able to extract any such substring in $\Oh{\ell + \log \log \log n'}$ time.  After \(\log^* n'\) iterations, we have stored $\Oh{\log^* n'}$ words and can extract any such substring in $\Oh{\ell}$ time.

\begin{lemma} \label{lem:bookmark}
Given a balanced SLP for a string \(S' [1..n']\) and a specified position in $S'$, we can add $\Oh{\log^* n'}$ words to the SLP such that, if a substring of length $\ell$ crosses that position, then we can extract that substring in $\Oh{\ell}$ time.
\end{lemma}

Applying Lemma~\ref{lem:bookmark} to each of the positions in $S'$ of the phrase boundaries in the LZ77 parse of $S$, then combining the resulting data structure with our instance of Bille et al.'s data structure for $S$, we obtain the following corollary.

\begin{corollary} \label{cor:extraction}
Given an SLP for $S$ with $r$ rules and a list of $t$ specified positions, we can store $S$ in $\Oh{r + t + z \log \log n}$ space such that, if a substring of length $\ell$ crosses a specified position, then we can extract that substring in $\Oh{\ell}$ time.
\end{corollary}

Applying Corollary~\ref{cor:extraction} to $S$ and choosing the $t$ specified positions to be the phrase boundaries in the LZ77 parse, we obtain a data structure that takes $\Oh{r + z \log \log n}$ space and supports extraction in $\Oh{\ell + \log n}$ time and extraction from around phrase boundaries in $\Oh{\ell}$ time.  Combining that with our modification of Kreft and Navarro's self-index from Section~\ref{sec:KN12}, we obtain our first main result.

\begin{theorem} \label{thm:unbalanced}
Given a straight-line program with $r$ rules for a string \(S [1..n]\) whose LZ77 parse consists of $z$ phrases, we can store a self-index for $S$ in $\Oh{r + z \log \log n}$ space such that we can extract any substring of length $\ell$ in $\Oh{\ell + \log n}$ time and, given a pattern \(P [1..m]\), we can list the $\occ$ occurrences of $P$ in $S$ in $\Oh{m^2 + \occ \log \log n}$ time.
\end{theorem}

We note that this self-index supports fast circular pattern matching (see, e.g.,~\cite{IR08}), for which we want to find all the cyclic shifts \(P [j + 1..m] P [1..j]\) of $P$ that occur in $S$.  Listing the occurrences can be handled in the same way as listing occurrences of $P$, so we ignore that subproblem here.  We modify our searching algorithm such that, when we would search in the first Patricia tree for the reverse \((P [1..i])^R\) of a prefix of $P$ and in the second Patricia tree for the corresponding suffix \(P [i + 1..m]\), we instead search for \((P [i + 1..m] P [1..i])^R\) and \(P [i + 1..m] P [1..i]\), respectively.  We record which nodes we visit in the Patricia trees and, when we stop descending (possibly because there is no edge whose label starts with the correct character), we extract the path label for the last node we visit in either tree and compute how many nodes' path labels match prefixes of \((P [i + 1..m] P [1..i])^R\) and \(P [i + 1..m] P [1..i]\).

For each node $v$ we visit in the first Patricia tree whose path label matches a prefix of \((P [i + 1..m] P [1..i])^R\), we find the first node $w$ (if one exists) that we visit in the second Patricia tree whose path label matches a prefix of \(P [i + 1..m] P [1..i]\) and such that the sum of the lengths of the path labels of $v$ and $w$ is at least $m$.  For each such pair, we perform a range-emptiness query (i.e., a range-reporting query that we stop early, determining only whether there are any points in the range) to check whether there are any phrase boundaries that are immediately preceded by the reverse of $v$'s path label and immediately followed by $w$'s path label.  These phrase boundaries are precisely those that are crossed by cyclic shifts of $P$ with the boundary between \(P [i]\) and \(P [i + 1]\).  This takes a total of $\Oh{m^2 \log \log z}$ time.

A similar idea works for finding the maximal substrings of $P$ that occur in $S$.  For each \(1 \leq i \leq m\), we can use doubling search --- with a range-emptiness query at each step --- to find the longest suffix \(P [h..i]\) of \(P [1..i]\) such that some phrase boundary is immediately preceded by \(P [h..i]\) and immediately followed by \(P [i + 1]\).  We then use doubling search to find the longest prefix \(P [i + 1..j]\) of \(P [i + 1..m]\) such that some phrase boundary is immediately preceded by \(P [h..i]\) and immediately followed by \(P [i + 1..j]\).  Notice that \(P [h..j]\) is the leftmost maximal substring of $P$ crossing a phrase boundary at position $i$, and we record it as a candidate maximal substring of $P$ occurring in $S$.

We now use doubling search to find the longest suffix \(P [h'..i]\) of \(P [h + 1..i]\) such that some phrase boundary is immediately preceded by \(P [h'..i]\) and immediately followed by \(P [i + 1..j + 1]\), then we use doubling search to find the longest prefix \(P [i + 1..j']\) such that some phrase boundary is immediately preceded by \(P [h'..i]\) and immediately followed by \(P [i + 1..j']\).  Notice that \(h' > h\), \(j' > j\) and \(P [h'..j']\) is the second maximal substring of $P$ crossing a phrase boundary at position $i$, and we record it as another candidate maximal substring of $P$ occurring in $S$.

We repeat this procedure until we have recorded all the candidate maximal substrings crossing a phrase boundary at position $i$.  While we work, the left endpoints of the prefixes and right endpoints of the suffixes we consider do not move left, so we use a total of $\Oh{m \log \log z}$ time to find the candidates associated with each position $i$.  Since two candidate associated with the same position cannot contain each other, there are at most $m$ of them.  Once we have all the candidates for every position $i$, finding the true maximal substrings of $P$ that occur in $S$ takes $\Oh{m^2}$ time.  In total we use $\Oh{m^2 \log \log z}$ time.

\begin{corollary} \label{cor:circular}
Given a pattern \(P [1..m]\), we can use the self-index described in Theorem~\ref{thm:unbalanced} to compute in $\Oh{m^2 \log \log z}$ time all the cyclic shifts and maximal substrings of $P$ that occur in $S$.
\end{corollary}

\section{Self-Indexing with a Balanced SLP} \label{sec:balanced}

In this section we describe how, if the SLP we are given for $S$ happens to be balanced, then we can improve the time bound in Theorem~\ref{thm:unbalanced} using Karp-Rabin hashes~\cite{KR87}.  A Karp-Rabin hash function
\[f (T [1..\ell]) = \left( \sum_{j = 1}^\ell \sigma^{\ell - j} T [j] \right) \bmod q\]
maps strings to numbers, where $\sigma$ is the size of the alphabet, $q$ is a prime and we interpret each character \(T [j]\) as a number between 0 and \(\sigma - 1\).  If we choose $q$ uniformly at randomly from among the primes at most $n^c$ then, for any two distinct strings \(T [1..\ell]\) and \(T' [1..\ell]\) with \(\ell \leq n\), the probability that \(f (T) = f (T')\) is $\Oh{c \log \sigma / n^{c - 1}}$.  Therefore, we can use Karp-Rabin hashes that fit in $\Oh{1}$ words with almost no chance of collisions.  Notice that, once we have computed the Karp-Rabin hashes of all the prefixes of a string, we can compute the Karp-Rabin hash of any substring in $\Oh{1}$ time.  Moreover, given the Karp-Rabin hashes of two strings, we can compute the Karp-Rabin hash of their concatenation in $\Oh{1}$ time.  We can replace Karp-Rabin hashing by deterministic alternatives~\cite{MSU97,ABR00} at the cost of increasing our bounds by polylogarithmic factors.

Consider the problem of finding the lexicographic range of the suffixes starting with \(P [i + 1..m]\) at phrase boundaries.  The problem of finding the lexicographic range of reversed phrases ending with \(P [1..i]\) is symmetric. Suppose we augment the Patricia tree for the suffixes by storing at each node $u$ the Karp-Rabin hash of $u$'s path label.  This takes $\Oh{z}$ extra space and, assuming our Karp-Rabin hash causes no collisions and we have already computed the Karp-Rabin hashes of all the prefixes of $P$, lets us find the deepest node $v$ whose path label is a prefix of \(P [i + 1..m]\) in time proportional to $v$'s depth.  In the worst case, however, $v$'s depth could be as large as \(m - i\).  Fortunately, while studying a related problem, Belazzougui, Boldi, Pagh and Vigna~\cite{BBPV09,BBPV10} showed how, by storing one Karp-Rabin hash for each edge, we can use a kind of binary search to find $v$ in $\Oh{\log m}$ time.  Ferragina~\cite{Fer11} gave a somewhat simpler solution in which he balanced the Patricia tree by a centroid decomposition.  His solution also takes $\Oh{z}$ space but with $\Oh{\log z}$ search time.

If the length of $v$'s path label is exactly \(m - i\) then, again assuming our Karp-Rabin hash causes no collisions, $v$'s path label is \(P [i + 1..m]\).  Otherwise, $v$'s path label is a proper prefix of \(P [i + 1..m]\) and in $\Oh{1}$ time we can find the edge descending from $v$ (if one exists) whose label begins with the next character of \(P [i + 1..m]\).  Let $w$ be the child of $v$ below this edge.  If the length of $w$'s path label is at most \(m - i\) then we know by our choice of $v$ that no suffix starts at a phrase boundary with \(P [i + 1..m]\).  Assume $w$'s path label has length at least \(m - i + 1\).  If any suffix starts at a phrase boundary with \(P [i + 1..m]\), then those that do correspond to the leaves in $w$'s subtree.  We cannot determine from looking at Karp-Rabin hashes stored in the Patricia tree, however, whether there are any such suffixes.  In order to determine this, we use the balanced SLP to compute the Karp-Rabin hash of the first \(m - i\) characters of $w$'s path label.

Recall from Section~\ref{sec:unbalanced} that, given a balanced SLP with $r$ rules for a string $S$, a specified position $i$ in $S$ and a bound $L$, we can add $\Oh{1}$ space such that later, given a length \(\ell \leq L\), in $\Oh{\log L}$ time we can compute a list of $\Oh{\log \ell}$ terminal and non-terminal symbols such that the concatenation of their expansions is \(S [i..i + \ell - 1]\).  (This is the same information we store to extract \(S [i..i + \ell - 1]\).)  It follows that, if we store the Karp-Rabin hash of the expansion of every non-terminal symbol, then we can compute the Karp-Rabin hash of \(S [i..i + \ell - 1]\) in $\Oh{\log L}$ time.  Symmetrically, we can add $\Oh{1}$ space such that we can compute in $\Oh{\log L}$ time the Karp-Rabin hash of any substring of length at most $L$ that ends at position $i$.  Therefore, we can add $\Oh{1}$ space such that we can compute in $\Oh{\log L}$ time the Karp-Rabin hash of any substring of length at most $L$ that crosses position $i$ in $S$.  As long as $L$ is polynomial in the length $\ell$ substring whose Karp-Rabin hash we want, \(\log L = \Oh{\log \ell}\).  If we fix \(\epsilon > 0\) and apply this construction with $L$ set to each of the $\Oh{\log \log n}$ values \(n^\epsilon, n^{\epsilon^2}, n^{\epsilon^3}, \ldots, 2\), then we obtain the following result.

\begin{lemma} \label{lem:hashes}
Given a balanced SLP for a string \(S [1..n]\) and a specified position in $S$, we can add $\Oh{\log \log n}$ words to the SLP such that, if a substring of length $\ell$ crosses that position, then we can compute its Karp-Rabin hash in $\Oh{\log \ell}$ time.
\end{lemma}

Applying Lemma~\ref{lem:hashes} to each of the phrase boundaries in the LZ77 parse of $S$, we obtain the following corollary.

\begin{corollary} \label{cor:hashes}
Given a balanced SLP for $S$ with $r$ rules, we can store $S$ in $\Oh{r + z \log \log n}$ space such that, if a substring of length $\ell$ crosses a phrase boundary, then we can compute its Karp-Rabin hash in $\Oh{\log \ell}$ time.
\end{corollary}

Combining Corollary~\ref{cor:hashes} with Belazzougui et al.'s construction, we obtain a data structure that takes $\Oh{r + z \log \log n}$ space and allows us to find in $\Oh{\log m}$ time the lexicographic range of the suffixes starting with \(P [i + 1..m]\) at phrase boundaries, assuming our Karp-Rabin hash causes no collisions and we have already computed the Karp-Rabin hashes of all the prefixes of $P$.  Since computing the Karp-Rabin hashes of all the prefixes of $P$ takes $\Oh{m}$ time and we need do it only once, it follows that we can find in a total of $\Oh{m \log m}$ time the lexicographic range of the suffixes starting with \(P [i + 1..m]\) for every value of $i$ and, symmetrically, the lexicographic range of the reversed prefixes ending with \(P [1..i]\).  Combining this data structure with Lemma~\ref{lem:bookmark}, we can also support extraction in $\Oh{\log n + \ell}$ time and extraction from around phrase boundaries in $\Oh{\ell}$ time.  Combining this data structure with our modification of Kreft and Navarro's self-index from Section~\ref{sec:KN12}, we obtain our second main result, below, except that our search time is $\Oh{m \log m + m \log \log z + \occ \log \log n}$ instead of $\Oh{m \log m + \occ \log \log n}$.

\begin{theorem} \label{thm:balanced}
Given a balanced straight-line program with $r$ rules for a string \(S [1..n]\) whose LZ77 parse consists of $z$ phrases, we can store a self-index for $S$ in $\Oh{r + z \log \log n}$ space such that we can extract any substring of length $\ell$ in $\Oh{\ell + \log n}$ time and, given a pattern \(P [1..m]\), we can list the $\occ$ occurrences of $P$ in $S$ in $\Oh{m \log m + \occ \log \log n}$ time.  Our construction is randomized but, given any constant $c$, we can bound by \(1 / n^c\) the probability that we build a faulty index.
\end{theorem}

Unfortunately, this time we cannot use Bille and G{\o}rtz'~\cite{BG11} approach alone to eliminate the $\Oh{m \log \log z}$ term.  When \(m \leq \log \log z\), storing one-dimensional range-reporting data structures at nodes in the top \(\log \log z\) levels of the Patricia trees means we use $\Oh{r + z \log \log n}$ space and $\Oh{m \log m + \occ \log \log n}$ search time; when \(m \geq \log z\), the $\Oh{m \log m}$ term dominates the $\Oh{m \log \log z}$ term anyway. To deal with the case \(\log \log z < m < \log z\), we build a Patricia tree for the set of $\Oh{z \log z}$ substrings of $S$ that cross a phrase boundary, start at most \(\log z\) characters before the first phrase boundary they cross, and end exactly \(\log z\) characters after it (or at \(S [n]\), whichever comes first).  At the leaf corresponding to each such substring, we store $\Oh{\log \log z}$ bits indicating the position in the substring where it first crosses a phrase boundary.  In total this Patrica tree takes $\Oh{z \log \log z}$ words.

If \(\log \log z < m < \log z\), we search for $P$ in this new Patricia tree, which takes $\Oh{m}$ time.  Suppose our search ends at node $v$.  If $P$ occurs in $S$, then the leaves in $P$'s subtree store the distinct positions in $P$'s primary occurrences where they cross phrase boundaries.  To determine whether $P$ occurs in $S$, it suffices for us to choose any one of those positions, say $i$, and check whether there is a phrase boundary immediately preceded by \(P [1..i]\) and immediately followed by \(P [i + 1..m]\).  To do this, we search in our first two augmented Patricia trees and perform a range-emptiness query.  If \(m \leq \log \log z\) time then we can perform the range-emptiness query with the one-dimensional range-reporting data structures in $\Oh{1}$ time; otherwise, we perform the range-emptiness query with our data structure for four-sided range reporting in \(\Oh{\log \log z} \subseteq \Oh{m}\) time.  If we learn that $P$ does not occur in $S$, then we stop here, having used a total of $\Oh{m}$ time.  If we learn that $P$ does occur in $S$, then in $\Oh{\occ}$ time we traverse $v$'s subtree to obtain the full list of distinct positions in $P$'s primary occurrences where they first cross phrase boundaries.  For each such position, we search in our first two augmented Patricia trees and perform a range-reporting query.  This takes $\Oh{m \log m + \occ \log \log z}$ time and gives us the positions of all $P$'s primary occurrences in $S$.

\section{Future Work} \label{sec:future}

We are currently working on a practical implementation of our self-index.  We believe the most promising avenue is to use Maruyama, Sakamoto and Takeda's~\cite{MST12} algorithm to build a balanced SLP, a wavelet tree as the range-reporting data structure~\cite{GGV03,Nav12} and Ferragina's~\cite{Fer11} restructuring to balance the Patricia trees.  When \(m \leq z\) --- which is the case of most interest for many applications in bioinformatics --- this implementation should take $\Oh{r}$ space and support location of all $\occ_1$ primary occurrences in $\Oh{(m + \occ_1) \log z}$ time, with reasonable coefficients.  As we have explained, finding all secondary occurrences is relatively easy once we have found all the primary occurrences.

Approximate pattern matching is often more useful than exact pattern matching, especially in bioinformatics.  Fortunately, Russo, Navarro, Oliveira and Morales~\cite{RNOM09} showed how to support practical approximate pattern matching using indexes for exact pattern matching, and we believe most of their techniques are applicable to our self-index.  One potential problem is how to perform backtracking using Patricia trees augmented with Karp-Rabin hashes, without storing or extracting edge labels.  This is because comparing hashes tells us (with high probability) when strings differ, but it does not tell us by how much they differ.  We are currently investigating a new variant of Karp-Rabin hashes by Policriti, Tomescu and Vezzi~\cite{PTV??} that roughly preserves Hamming distance.

Finally, we have shown elsewhere~\cite{GGP11} that supporting extraction from specified positions has applications to, e.g., sequential approximate pattern matching.  In that paper we developed a different data structure to support such extraction, which we have now implemented and found to be faster and more space-efficient than Kreft and Navarro's solutions.  Nevertheless, we expect the solutions we have given here to be even better.

\section*{Acknowledgments}

Many thanks to Djamal Belazzougui, Francisco Claude, Veli M\"akinen, Gonzalo Navarro and Jorma Tarhio, for helpful discussions.

\bibliographystyle{model1-num-names}
\bibliography{faster}

\end{document}